# Laboratory Modeling of Supernova Remnants Collisions: Implications for Triggered Star Formation

Marin Fontaine, Clotilde Busschaert, and Émeric Falize
CEA, DAM, DIF, F-91297 Arpajon, France; marin.fontaine@cea.fr

## Abstract

Theoretical models of star formation consistently underestimate the rates observed in astronomical surveys. Stars form within giant molecular clouds, which fragment into dense clumps under the combined influences of turbulence, magnetic fields, radiation, and gravity. While some of these clumps collapse spontaneously, others require an external trigger, a mechanism estimated to account for 14%–25% of star formation in regions such as the Elephant Trunk Nebula. Laboratory astrophysics has emerged as a powerful approach for investigating such triggering processes, in particular those involving supernova remnants (SNRs). Recent experiments, guided by well-established scaling laws, have successfully replicated the dynamics of SNRs and their interactions with dense clumps or other SNRs. In this work, we present a comprehensive numerical study of these experimental configurations using the 3D radiation-hydrodynamics code TROLL. The simulations provide enhanced insight into the underlying physical mechanisms, accurately reproduce key experimental phenomena, and offer valuable comparisons with analytical models. This study underscores the strong synergy between laboratory experiments and numerical simulations, laying a robust foundation for future advancements in laboratory astrophysics. Furthermore, we propose a new experimental setup that offers improved scaling for the asymmetric collision observed in the DEM L316 system. Our findings also show that SNR collisions in dense environments can decrease the gravitational stability of dense clumps, thereby promoting their collapse and potentially triggering star formation.

## 1. Introduction

New stars form within giant molecular clouds (GMCs) through a complex interplay of physical processes (C. F. McKee & E. C. Ostriker 2007). These massive gas structures fragment into dense clumps under the combined influence of gravity (E. Jaupart & G. Chabrier 2020), magnetic fields (C. Federrath & R. S. Klessen 2012; J. D. Soler & P. Hennebelle 2017), radiation (C. de Boisanger & J. P. Chieze 1991; R. M. Jennings & Y. Li 2021), and turbulence (P. Hennebelle et al. 2024). The fate of these clumps depends on their stability; while some remain in quasi-equilibrium, others undergo gravitational collapse, giving rise to protostars. Despite considerable progress in modeling efforts, theoretical frameworks continue to underestimate the observed star formation rates (SFRs; M. R. Krumholz 2014). One proposed explanation for this discrepancy involves external triggering, wherein nearby astrophysical events, such as protostellar outflows or supernova (SN) explosions, initiate or accelerate the collapse of otherwise stable clumps. In IC 1396A, commonly referred to as the Elephant Trunk Nebula, such externally driven mechanisms are estimated to contribute 14%–25% of the total SFR (K. V. Getman et al. 2012). This phenomenon has been studied across various astrophysical contexts, including protostellar jets (P. C. Fragile et al. 2017; M. Fontaine et al. 2025), cloud–cloud collisions (Y. Fukui et al. 2021), and supernova remnants (SNRs; M. Völschow et al. 2017).

Within the context of SNRs, several mechanisms have been proposed by which star formation may be externally triggered. In dense environments, SNR ejecta can impact nearby stable clumps, compressing them via transmitted shock waves. This compression may initiate gravitational collapse, potentially resulting in the formation of new stars. Observations using instruments such as ALMA, targeting regions in the Large Magellanic Cloud, have revealed heating and enhanced emission in GMCs impacted by nearby SNRs, providing compelling evidence for this triggering mechanism (H. Sano et al. 2023).

Another plausible pathway involves direct, head-on collisions between two SNRs. These interactions produce a region of intense physical alteration, where conditions differ markedly from the surrounding medium. Numerical simulations suggest that densities in these collision regions can increase by up to a factor of four, accompanied by elevated temperatures and emission signatures (J. C. Toledo-Roy et al. 2009). The DEM L316 system has drawn significant attention, as it may represent a genuine SNR–SNR collision rather than a simple line-of-sight projection (M. Nishiuchi et al. 2001; R. M. Williams & Y. H. Chu 2005). Regardless of its exact nature, both simulations and observations indicate that such interaction zones can develop substantial vorticity. This vorticity may drive magnetic field generation through the Biermann battery mechanism (K. M. Schoeffler et al. 2016). Increased turbulence and magnetization within the medium could, in turn, enhance cloud fragmentation, thereby fostering conditions favorable to star formation.

Advances in experimental techniques have enabled the reproduction of astrophysical processes in laboratory environments, in particular through the use of high-energy lasers and pulsed power facilities (B. A. Remington et al. 2006; S. Lebedev et al. 2019). These approaches allow for controlled and scalable experiments, making it possible to investigate the complex dynamics of SNRs under well-defined conditions (P. Mabey et al. 2020; A. Triantafyllidis et al. 2025). Among the various processes studied, particular attention has been given to external triggering mechanisms of star formation involving SNRs. Two key configurations have been explored experimentally in the context of star formation triggering by SNRs: the collision between two expanding SNRs (B. Albertazzi et al. 2020) and the impact of an SNR shock front on a dense clump embedded in its surrounding medium (B. Albertazzi et al. 2022).

This study presents a numerical reproduction of both experimental scenarios. Three-dimensional radiation-hydrodynamics simulations were performed using the ALE radiation-hydrodynamics code TROLL (E. Lefebvre et al. 2018), which has been specifically adapted to emulate the conditions of laser-driven experiments. These simulations, once the TROLL code's ability to reproduce the experimental results has been demonstrated, offer access to critical physical parameters that cannot be directly measured through experimental diagnostics, thereby providing complementary insights. The results facilitate direct comparison with analytical models and previous experimental and numerical studies, ultimately deepening our understanding of the governing physical mechanisms.

Section 2 outlines the theoretical background for SNR–SNR and SNR–clump collisions, and discusses the scaling relationships connecting astrophysical and laboratory regimes. Section 3 describes the numerical setup in detail, including the experimental configurations and simulation tools used. Finally, Section 4 presents the numerical results on SNR-driven external triggering of star formation, comparing them with analytical models and offering an astrophysical interpretation.

## 2. Theoretical Description

The evolution of an SNR and the expansion of its ejecta are well-characterized by theoretical models. This process unfolds through four principal stages: the ballistic phase immediately following the explosion, the Sedov phase, the snowplow phase, and ultimately, the remnant's integration into the interstellar medium (T. Padmanabhan 2001; E. R. Micelotta et al. 2018). The present study focuses on the first two phases, which typically span from 10,000 to 100,000 yr.

The ballistic, or free-expansion, phase persists for several hundred years and is defined by a constant ejecta velocity, $v_{\rm ej}$, as it propagates into a uniform interstellar medium of low density, $\rho_0$. When the accumulated interstellar mass becomes comparable to the ejecta mass, $M_{\rm ej}$, the remnant begins to decelerate and transitions into the Sedov phase. The time at which this transition occurs, denoted $t_{\rm ST}$, can be estimated as (J. P. Ostriker & C. F. McKee 1988)

$$t_{\rm ST} = \frac{R_{\rm ST}}{v_{\rm ej}} = \left(\frac{3M_{\rm ej}}{4\pi\rho_0 v_{\rm ej}^3}\right)^{1/3}. \quad (1)$$

Here, $R_{\rm ST}$ corresponds to the radius at which the transition into the Sedov–Taylor phase takes place. During the Sedov phase, the shock front propagates as a self-similar blast wave (G. I. Barenblatt 1996), described by the Sedov–Taylor solution, wherein the shock radius $R_s$ evolves according to

$$R_s(t) = \left(\xi_0 \frac{E_0}{\rho_0}\right)^{1/5} t^{2/5}, \quad (2)$$

where $\xi_0$, $E_0$, and $t$ represent the self-similar constant, the explosion energy, and time, respectively. Drawing from the analytical formulation of blast waves by G. G. Bach & J. H. S. Lee (1970), and under the assumption of a strong shock regime ($M^2 \gg 1$, with $M = v_{\rm ej}/c_0$ denoting the Mach number and $c_0$ denoting the sound speed within the ambient medium), the spatial and temporal profiles for velocity $v(r, t)$, density $\rho(r, t)$, and pressure $P(r, t)$ are derived as follows:

$$v(r, t) = \frac{2}{5(\gamma+1)}\frac{r}{t}\left[1 + 3\frac{\gamma-1}{\gamma+1}\ln\left(\frac{r}{R_s(t)}\right)\right],$$

$$\rho(r, t) = \rho_0\frac{\gamma+1}{\gamma-1}\left(\frac{r}{R_s(t)}\right)^q,$$

$$P(r, t) = \rho_0 \dot{R}_s^{\,2}\left[\frac{2}{\gamma+1} + \frac{1}{j+\gamma}\left(\left(\frac{r}{R_s(t)}\right)^{q+2} - 1\right)\left(1 - \frac{2}{\gamma+1} + \frac{3}{2t^2}\right)\right]. \quad (3)$$

Here, $\gamma$ is the adiabatic index, $r$ the radial position, $q = 2(j+1)/(\gamma-1)$ an exponent, and $j$ the geometrical factor, where $j = 0$ corresponds to one-dimensional symmetry, $j = 1$ to two-dimensional symmetry, and $j = 2$ to three-dimensional symmetry.

These expressions describe the temporal evolution of blast wave profiles during the Sedov phase, provided the shock retains a high Mach number. As will be demonstrated later, this condition remains valid over the majority of the experimental and simulation timescales considered. It is within this phase that the SNR reaches scales of several parsecs, allowing it to interact with surrounding astrophysical structures, such as dense clumps within molecular clouds or nearby SNRs. These interactions can be effectively modeled using the blast wave framework introduced above.

A more rigorous theoretical treatment of blast waves in astrophysical contexts was presented by J. K. Truelove & C. F. McKee (1999), who established the power-law structure of nonradiative SNRs and validated these results through comparison with astronomical observations. Their framework also incorporates the transition between the ballistic and Sedov–Taylor phases into the governing equations for the temporal evolution of the shock radius and velocity.

### 2.1. Symmetrical Collision of Two SNRs

Sedov–Taylor blast waves are supersonic, collisional shocks. A symmetric collision between two counter-propagating blast waves is, from a hydrodynamic perspective, equivalent to the reflection of a single blast wave off a rigid boundary. This setup leads to the formation of a distinct interaction region (see Figure 1).

Under the strong shock approximation ($M = V_s/c_0 \gg 1$), the properties at the shock front of a single blast wave can be described using analytical relations derived from the Rankine–

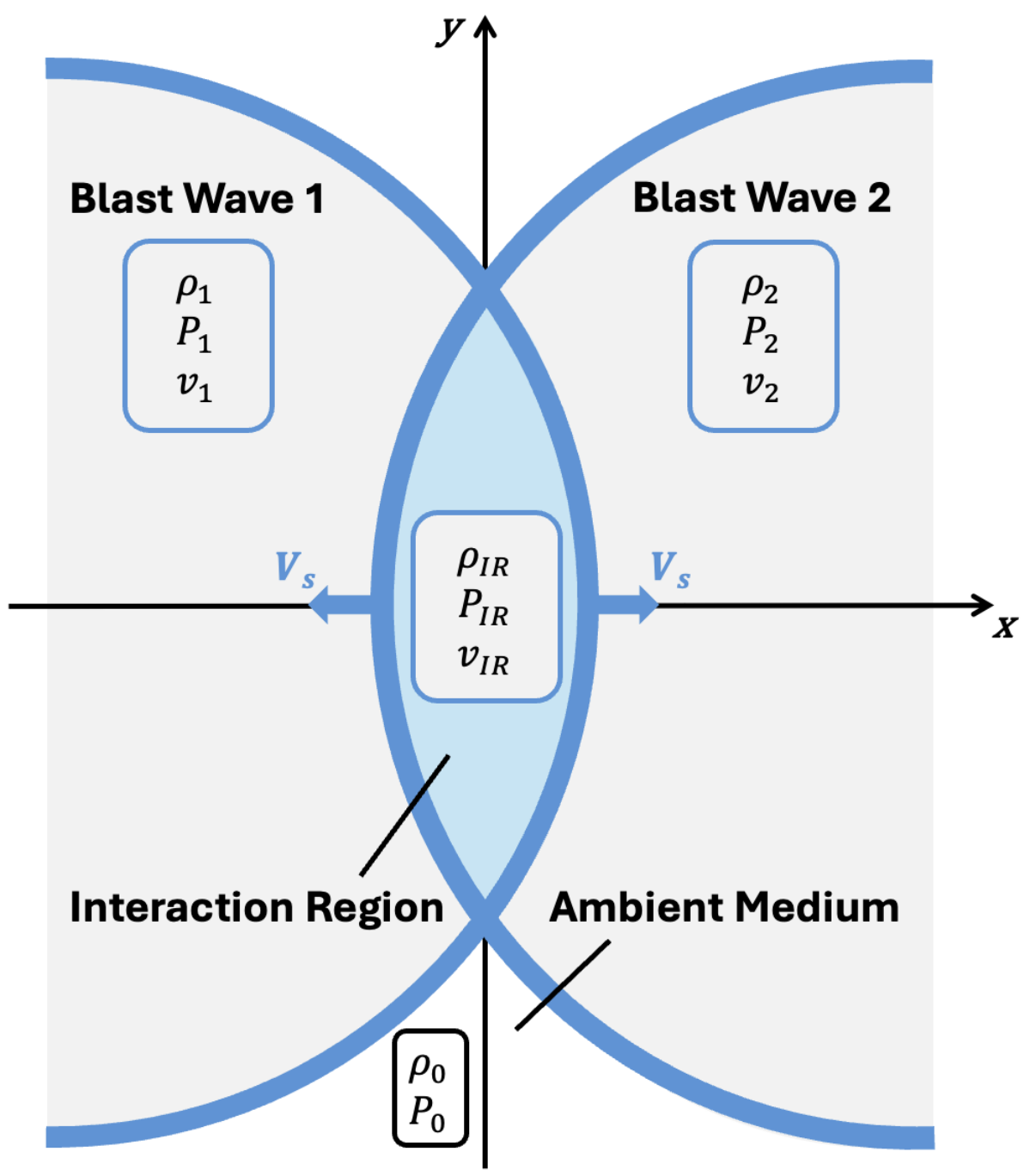

**Figure 1.** Schematic representation of the collision between two blast waves, resulting in the formation of an interaction region.

Hugoniot conditions:

$$\rho_1 = \frac{\gamma+1}{\gamma-1}\rho_0, \quad P_1 = \frac{2}{\gamma+1}\rho_0 v_1^2, \quad v_1 = \frac{\gamma-1}{\gamma+1}v_0. \quad (4)$$

Here, $\rho_1$, $P_1$, and $v_1$ represent the density, pressure, and velocity of the shocked material within the first blast wave, while $\rho_0$, $P_0$, and $v_0$ denote the corresponding properties of the undisturbed ambient medium (as shown in Figure 1). These relations are derived from Equations (3) by evaluating the limit as $r = R_s$, and therefore do not describe the internal structure of the blast wave. In a symmetrical collision, both blast waves arrive with identical properties behind the shock front. The parameters of the second blast wave can be expressed as $\rho_2 = \rho_1$, $P_2 = P_1$, and $v_2 = -v_1$, reflecting the opposing propagation directions of the shock fronts.

At the moment of collision, in one dimension, the velocities of the incoming blast waves are preserved, and the resulting central interaction region acquires the properties $\rho_{\rm IR}$, $P_{\rm IR}$, and $v_{\rm IR}$ (R. Y. Tugazakov & A. S. Fonarev 1974):

$$\rho_{\rm IR} = \frac{\gamma+1}{\gamma-1}\rho_0, \quad P_{\rm IR} = 2P_1 = \frac{4u_1^2\rho_0}{\gamma+1}, \quad v_{\rm IR} = 0. \quad (5)$$

One-dimensional numerical simulations further describe the temporal evolution of density, pressure and velocity within this region (R. Y. Tugazakov & A. S. Fonarev 1974). In an additional study, H. L. Brode (1977) examined the peak overpressure produced by the collision of two simultaneous blast waves. A key distinction in his work is the inclusion of the spherical geometry of the counter-propagating waves. This geometry leads to a higher compression caused by the angle at which the shock fronts meet (see Figure 1). His analysis revealed that the central pressure can greatly exceed one-dimensional predictions, with a peak ratio of $P_{\rm IR}/P_1 \sim 5$–7. These findings underscore the significant role of the three-dimensional collision in amplifying compression during such collisions.

### 2.2. Influence of a Delay Time on the Collision

When the two SNe do not occur simultaneously, the resulting blast waves are no longer symmetric. This asymmetrical configuration has been explored both analytically and numerically, revealing notable differences in the structure and evolution of the interaction region compared to the symmetrical case. When the stronger shock wave, originating from the younger SNR, collides with the weaker one from the older SNR, the interaction region becomes divided by a moving contact discontinuity (H. L. Brode 1977). This discontinuity separates two regions characterized by different densities and temperatures, while maintaining equal velocities and pressures on both sides. The peak overpressure within the interaction region can be estimated analytically using the Rankine–Hugoniot relations for an ideal gas. In the H. L. Brode (1977) configuration, it typically reaches three to four times the pressure of the material behind the stronger shock, considerably lower than the peak pressures produced in collisions involving two shocks of equal strength. This scenario has been the subject of extensive numerical investigation (M. K. Almustafa & M. L. Nehdi 2023).

### 2.3. Impact of a Blast Wave on a Spherical Obstacle

When a strong blast wave encounters an obstacle, it produces both transmitted and reflected shocks, and typically forms a bow shock upstream of the obstacle (C. F. McKee & L. L. Cowie 1975). In the specific case where the obstacle is spherical, the pressure of the shocked material within the clump, denoted as $P_c$, can be expressed as

$$P_c = \frac{2}{\gamma_c+1}\rho_{0,c}v_c^2, \quad (6)$$

where $\gamma_c$, $\rho_{0,c}$, and $v_c$ correspond to the adiabatic index of the clump, its unshocked density, and the velocity of the transmitted shock, respectively. This velocity simplifies to

$$v_c \approx \frac{V_s}{\sqrt{\eta}}, \quad (7)$$

where $V_s$ is the velocity of the incoming blast wave shock front and $\eta = \rho_c/\rho_0$ represents the density contrast between the clump and the ambient medium. Comparable results were derived by R. A. Chevalier (1999) in the context of a radiatively cooling SNR propagating through a GMC with a radiative precursor.

These analytical expressions will serve as a valuable reference for interpreting the numerical results and assessing the compression of dense clumps impacted by SNR shocks.

### 2.4. Similarity Properties

To investigate the behavior of SNRs and their interactions in a controlled environment, it is essential to establish appropriate scaling laws. These laws ensure dynamic similarity between astrophysical phenomena and their laboratory counterparts, providing a rigorous mathematical framework for replicating such processes experimentally (D. Ryutov et al. 1999). When correctly derived, scaling laws allow for meaningful comparisons across theoretical models, experimental results, and

**Table 1**
Physical Properties of the Two SNRs Forming DEM L316, and the Laboratory Blast Wave

| Parameters | Shell A | Shell B | Scaled Shell A | Scaled Shell B | Laboratory Blast Wave |
|---|---|---|---|---|---|
| $r$ (cm) | $4.6 \times 10^{19}$ | $6.8 \times 10^{19}$ | 1 | 1 | 1 |
| $\rho$ (g cm$^{-3}$) | $1 \times 10^{-24}$ | $9 \times 10^{-25}$ | $4 \times 10^{-5}$ | $4 \times 10^{-5}$ | $4 \times 10^{-5}$ |
| $P$ (dyn cm$^{-2}$) | $1.4 \times 10^{-9}$ | $6.8 \times 10^{-10}$ | $1 \times 10^{8}$ | $1 \times 10^{8}$ | $1 \times 10^{8}$ |
| $t$ (s) | $9.5 \times 10^{11}$ | $1.3 \times 10^{12}$ | $4.9 \times 10^{-7}$ | $3.4 \times 10^{-7}$ | $1.5 \times 10^{-7}$ |
| $v$ (km s$^{-1}$) | 220 | 220 | 8.8 | 13 | 65 |
| Composition | H | H | ⋯ | ⋯ | $N_2$ |
| $T$ (eV) | 1400 | 1100 | ⋯ | ⋯ | 5 |
| Eu | 2.9 | 1.6 | 0.3 | 0.1 | 0.06 |
| $M$ | ∼20 | ∼20 | > 1 | > 1 | 6 − 10 |
| $\eta$ | $10^3$–$10^4$ | $10^3$–$10^4$ | $10^4$ | $10^4$ | $10^4$ |
| $\chi$ | ≫1 | > 1 | > 1 | > 1 | > 1 |

**Note.** The values for Shells A and B (corresponding columns) are taken from M. Nishiuchi et al. (2001) and R. M. Williams & Y. H. Chu (2005). Their scaling to the laboratory is indicated in the "Scaled Shell A" and "Scaled Shell B" columns. The values for the laboratory blast wave (corresponding columns) correspond to the time of collision.

numerical simulations, thereby reinforcing the credibility and applicability of laboratory astrophysics. The physical properties of the two SNRs in the DEM L316 system, along with those of the laboratory-generated blast wave, are summarized in Table 1.

Throughout the ballistic and Sedov phases, the ejecta evolves as a collisional, self-similar shock governed by the hydrodynamic equations. These shocks are considered nonradiative (J. K. Truelove & C. F. McKee 1999), as confirmed by the large cooling parameters $\chi$ listed in Table 1, defined as the ratio of the cooling time to the dynamical time. Consequently, the appropriate scaling must conserve the Euler equations for a polytropic, nonradiative flow. Following the methodology outlined in D. Ryutov et al. (1999), the radius $r$, density $\rho$, and pressure $P$ are scaled using constant factors $a$, $b$, and $c$, respectively, according to the following transformations:

$$r' = ar, \quad \rho' = b\rho, \quad P' = cP, \tag{8}$$

where each variable $x$ corresponds to the astrophysical scale and $x'$ to the laboratory scale. These transformations lead to the following constraints on time $t$ and velocity $v$:

$$t' = a\sqrt{\frac{b}{c}}\,t, \quad v' = \sqrt{\frac{c}{b}}\,v. \tag{9}$$

In this study, we aim to scale the DEM L316 system, comprising two SNRs of distinct ages, down to laboratory dimensions. This approach allows us to explore the similarity conditions for both SNRs. When applied to Shell A of DEM L316, the resulting scaling parameters are $a \sim 2.2 \times 10^{-20}$, $b \sim 4.0 \times 10^{19}$, and $c \sim 7.1 \times 10^{16}$. These yield a laboratory-equivalent SNR age of approximately 490 ns and a shock front velocity of about 8.8 km s$^{-1}$. For Shell B, the corresponding parameters are $a \sim 1.5 \times 10^{-20}$, $b \sim 4.4 \times 10^{19}$, and $c \sim 1.5 \times 10^{17}$, resulting in a laboratory age of roughly 340 ns and a shock velocity of approximately 13 km s$^{-1}$. These scaled values are detailed in the "Scaled Shell A" and "Scaled Shell B" columns of Table 1. Since both SNRs are in the same hydrodynamic regime, the scaling is individually valid in each case. Accordingly, the simulations in this work adopt scaling parameters that represent the average of the values obtained for each SNR. We note, however, that a rigorous hydrodynamic scaling would require a single set of parameters connecting an SNR to its laboratory counterpart, and that the averaged parameters used in this work do not theoretically describe each SNR individually, but rather provide the best possible representation of the system as a whole.

In addition, relevant dimensionless hydrodynamic numbers must be evaluated in both the astrophysical and laboratory regimes. The scaling preserves the Euler number, Eu $=P/\rho v^2$, ensuring consistent fluid behavior across scales. All considered blast waves are supersonic shocks, characterized by Mach numbers $M = V_s/c_0 \sim 10$. Another critical parameter is the density contrast, $\eta = \rho_c/\rho_b$, where $\rho_c$ represents the density of dense clumps and $\rho_b$ the density behind the SNR shock front. In GMCs, clump densities typically range from $10^4$ to $10^6$ cm$^{-3}$, which corresponds to the values of $\eta$ reported in Table 1 (J. P. Williams et al. 2000; J. Kauffmann et al. 2010). In the experimental setup, the density of the sphere was chosen specifically to reproduce this contrast.

## 3. Numerical Setup

This study presents the results of a numerical investigation conducted at the laboratory scale. The primary objective was to numerically reproduce the laboratory astrophysics experiments performed by B. Albertazzi et al. (2020, 2022) at the LULI2000 laser facility. Simulations were carried out using two distinct configurations, each designed to emulate the laboratory-scale generation of SNRs in accordance with established scaling laws. The following section outlines the numerical tools employed to construct the simulation framework, followed by a detailed description of the specific setups implemented.

All simulations were conducted using TROLL, the CEA's 3D radiation-hydrodynamics code. This Arbitrary Lagrange–Eulerian (ALE), multi-material code is specifically optimized for modeling the dynamics of laser-driven targets at laboratory scales. A full account of the numerical schemes used in TROLL is provided in E. Lefebvre et al. (2018). The laser pulse is discretized into numerical rays that are traced through the plasma, accurately modeling reflection, refraction, and absorption. Material properties, including the equation of state, opacity, and emissivity, are computed assuming local thermodynamic equilibrium, using dedicated codes and stored in tabulated form. This enables TROLL to simulate the full temporal evolution of the experiment, from laser irradiation through to the expansion of the resulting blast wave in the surrounding medium. Simulations were executed using up to several thousand processors, involving grids ranging from 1 to $7 \times 10^6$ cells, and requiring up to $2 \times 10^6$ CPU hours. The carbon pin surface was meshed with cells as thin as 10 nm to accurately resolve the laser–matter interaction. A mesh convergence study was performed to assess the sensitivity of the results to cell resolution, in particular with regard to laser energy absorption and subsequent blast wave development. To improve comparison with experimental diagnostics, X-ray radiography post-processing simulations were performed using the DIANE code (M. Caillaud et al. 2014; B. Avez et al. 2025). DIANE applies ray-tracing techniques to simulate the interaction between an X-ray source and the target, accounting for its geometry and hydrodynamic state. For consistency with the experimental setup, a monochromatic spectrum corresponding to the titanium He-$\alpha$ line at approximately 4.7 keV was used in the simulations.

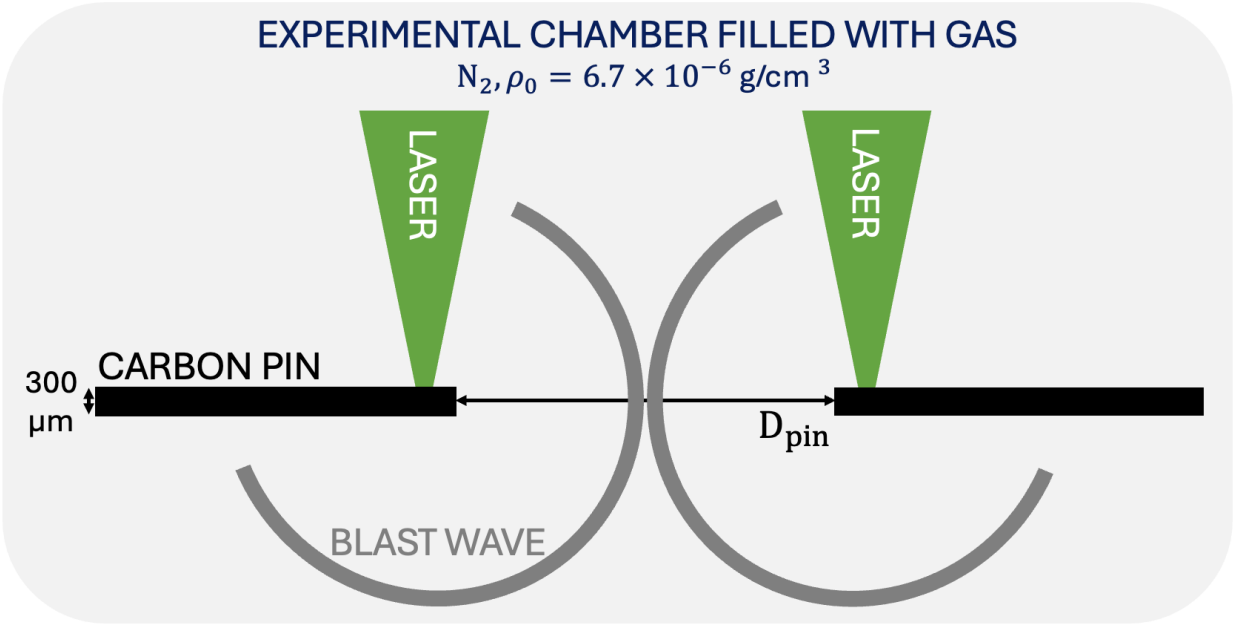

**Figure 2.** Numerical setup replicating the LULI2000 experiments on the collision of two blast waves. The distance between the pins, $D_{\rm pin}$, is set to either 1 or 2 cm.

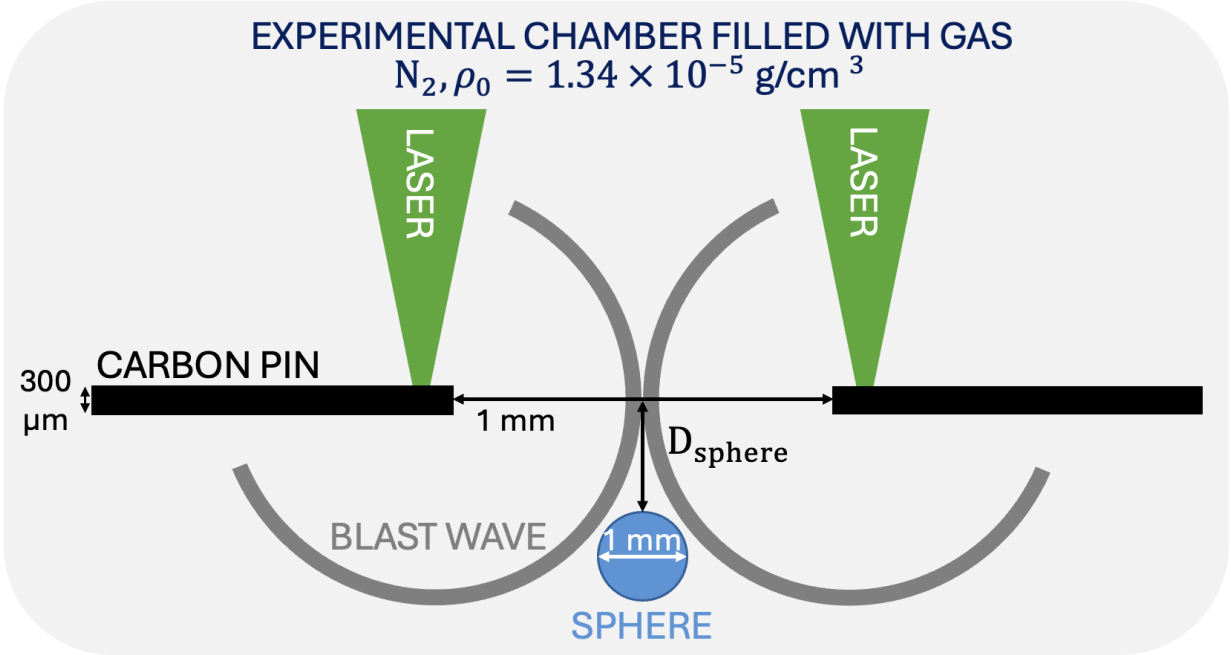

**Figure 3.** Numerical setup replicating the LULI2000 experiments on the collision of blast waves with spherical obstacles. The sphere's offset from the symmetry axis, $D_{\rm sphere}$, is set to 5, 2.5, or −0.5 mm.

To recreate an SNR in the laboratory, a solid carbon pin is placed within a nitrogen-filled chamber and irradiated with a 500 J laser pulse of 1 ns duration. This intense energy deposition induces rapid expansion of the pin material, launching a shock wave into the surrounding gas. The resulting dynamics reproduce the ballistic and Sedov–Taylor phases of an SNR under controlled conditions. The surrounding gas is uniform and no turbulence or instabilities develop on the experiment's timescale that could interfere with the blast wave's expansion and alter its morphology. This capability forms the foundation of the two experimental configurations explored in this study.

The first setup aims to generate two blast waves and examine their collision, along with the structure of the resulting interaction region. Following the experimental design detailed in Section II of B. Albertazzi et al. (2020), two symmetrically positioned carbon pins, separated by $D_{\rm pin} = 1$ cm, are each irradiated with a 500 J, $2\omega$, 1 ns laser pulse with a square temporal profile (see Figure 2). The laser is focused into a 200 $\mu$m diameter super-Gaussian spot, achieving an intensity of approximately $1 \times 10^{15}$ W cm$^{-2}$. The collision of the expanding blast waves is visualized using 2D schlieren diagnostics, which map gradients in electron density, providing a clear depiction of the shock fronts and valuable insight into the evolution of the interaction zone. This diagnostic was reproduced numerically from the 3D simulations via post-processed maps of the electronic density gradient, following 2D slices along directions of interest.

To better emulate the DEM L316 system, the pin separation was increased to $D_{\rm pin} = 2$ cm, approximately the sum of both their radii at the laboratory scale, and a temporal delay was introduced between the two laser pulses. Based on the scaling laws derived earlier, a 100 ns age difference between the two SNRs corresponds to the laboratory timescale. Accordingly, simulations were performed with delays of 75, 100, 125, and 150 ns between the pulses, enabling the study of asymmetrical SNR collisions.

The second setup, shown in Figure 3, focuses on investigating the interaction between blast waves and a dense spherical clump. In the experiment described in Sec III of B. Albertazzi et al. (2022), a 1 mm diameter CHO sphere with a density of 150 mg cm$^{-3}$ is placed at varying distances $D_{\rm sphere} = 0$, 2.5, and 5 mm off-axis from the carbon pins, which remain separated by $D_{\rm pin} = 1$ cm. In addition to schlieren imaging, X-ray radiography is employed to probe the sphere's internal structure and quantify its compression during the interaction.

After validating the simulations against the experimental data, the numerical setup was extended to examine how the clump's spatial position affects its interaction with the shock fronts. In this parametric study, the clump was placed either directly along the axis connecting the pins or offset by 2.5 mm.

## 4. Results and Numerical Study

Before presenting the 3D simulations of SNR collisions at laboratory scale, it is crucial to first characterize the resulting shock wave and verify that it exhibits the properties of a blast wave. Furthermore, the ability of the TROLL code to accurately replicate a blast wave driven by a laser pulse must be assessed.

### 4.1. 3D Simulation of a Single Blast Wave

Figure 4 presents the results of a 3D simulation of a single blast wave, conducted using the numerical setup described previously. Three snapshots at 50, 100, and 200 ns illustrate the temporal evolution of the shock front and its morphology. Employing a three-dimensional simulation is essential because the laser energy deposition (indicated by the green arrow in the first snapshot) is highly anisotropic and occurs off the cylindrical axis. Consequently, the blast wave expands preferentially in the direction of the laser pulse, with less expansion behind the pin. Despite this anisotropy, the small diameter of the pin ensures that the overall morphology remains an approximate spherical blast wave. The blast wave shell features a dense shock front enclosing a lower-density interior, characteristic of both laboratory blast waves and astrophysical SNRs.

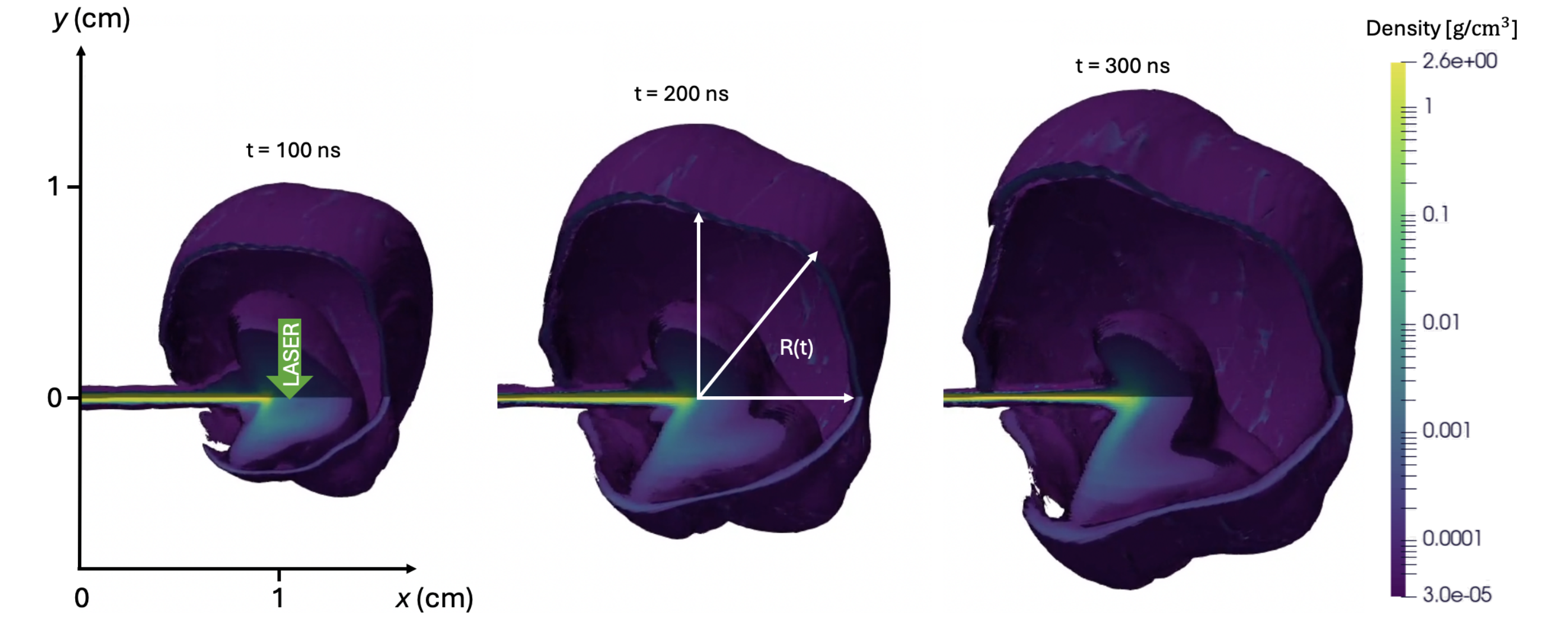

**Figure 4.** Snapshots of the 3D simulation at 50, 100, and 200 ns showing the density on a logarithmic scale. A quarter of the simulation domain has been removed along the $x$-axis, to reveal the internal structure of the expanding shell. Densities below $\rho = 3 \times 10^{-5}$ g cm$^{-3}$ were masked to enhance contrast. The green arrow in the first snapshot indicates the location of the laser pulse impact. White arrows labeled $R(t)$ denote the directions along which the expansion is analyzed.

To validate the laboratory reproduction of a Sedov–Taylor blast wave, several key physical quantities derived from the simulation are analyzed. According to the theoretical framework outlined previously, the shock front radius is expected to expand following Equation (2), implying a shock velocity evolving as $V_s \propto t^{-3/5}$. From the 3D simulation shown in Figure 4, the temporal evolution of the radius was extracted and compared to this theoretical power-law trend. The expansion was measured along three directions: the $x$-axis, the diagonal above the pin, and the direction of the laser pulse, as indicated by the white arrows in Figure 4. These results, covering the first 300 ns of expansion, are displayed in Figure 5.

The simulation profiles reveal a key feature observed experimentally: as in astrophysical SNRs, the blast wave initially undergoes a ballistic phase, during which the shock propagates at nearly constant velocity and does not follow the Sedov–Taylor relation. In this setup, and for the given initial energy, the ballistic phase lasts approximately 10 ns, reaching a radius of about 5 mm. The transition from this ballistic regime to the Sedov–Taylor phase occurs simultaneously in both simulation and experiment, consistent with the analytical estimate provided in Equation (1). Beyond this transition, the shock front radius evolves closely in accordance with theoretical predictions, exhibiting the expected power-law behavior. The anisotropic expansion of the blast wave is manifested as an increase of $E_0$ in Equation (2) and accounts for the faster growth observed in certain measured directions. The radius reached at 100 ns permits the calculation of $E_0$ values of approximately 20 J in the laser pulse direction, 4 J along the diagonal, and 1 J along the $x$-axis. When all directions are integrated, the initial blast wave energy approaches 10% of the laser energy, which is the amount measured experimentally (see Section III A of B. Albertazzi et al. 2020) and predicted theoretically (P. Mora 1982).

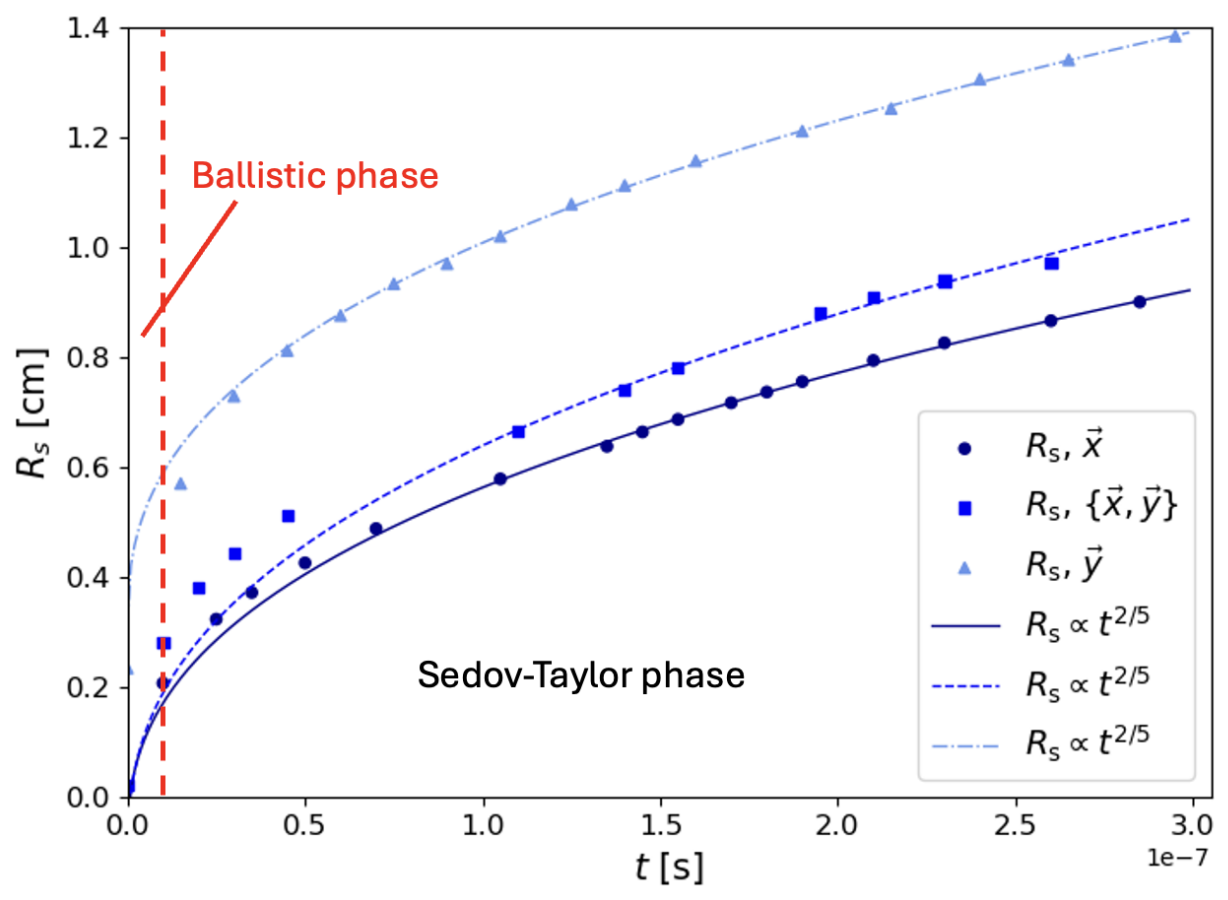

**Figure 5.** Temporal evolution of the shock front radius measured in three directions (see Figure 4), compared with the theoretical power-law behavior (solid lines). Measurements were taken along the $x$-axis (dark blue dots), the diagonal above the pin (blue squares), and the laser pulse direction (light blue triangles). The transition to the Sedov–Taylor phase, occurring around 10 ns, is indicated by the red dashed line.

The laboratory blast wave behaves as a strong shock characterized by a high Mach number, which decreases over time as the wave decelerates, reaching approximately 6 at 300 ns. Accordingly, the compression of the ambient gas may be estimated using the Rankine–Hugoniot jump conditions (Equation (4)), assuming nitrogen behaves as a diatomic ideal gas with adiabatic index $\gamma = 7/5$. Immediately behind the shock front, the density attains roughly six times the initial value of the surrounding gas, $\rho_1 = 6\rho_0 \approx 4 \times 10^{-5}$ g cm$^{-3}$, which is obtained both experimentally (see Section III B of B. Albertazzi et al. 2020) and numerically. This corresponds to a post-shock pressure of $P_1 \approx 1.5 \times 10^8$ dyn cm$^{-2}$, while the pre-shock value is 5700 dyn cm$^{-2}$. Finally, the pre-shock temperature was set to 1 eV in the simulations and goes up to approximately 5 eV behind the shock front along the $x$-axis, which is coherent with experimental measurements (see Section III D of B. Albertazzi et al. 2020). At this temperature, nitrogen is only weakly dissociated, given that the required energy is approximately 9.8 eV (see, for example, Figure 5 of S. Kodama et al. 2013). This low degree of dissociation supports the use of a diatomic description for the blast waves in this study.

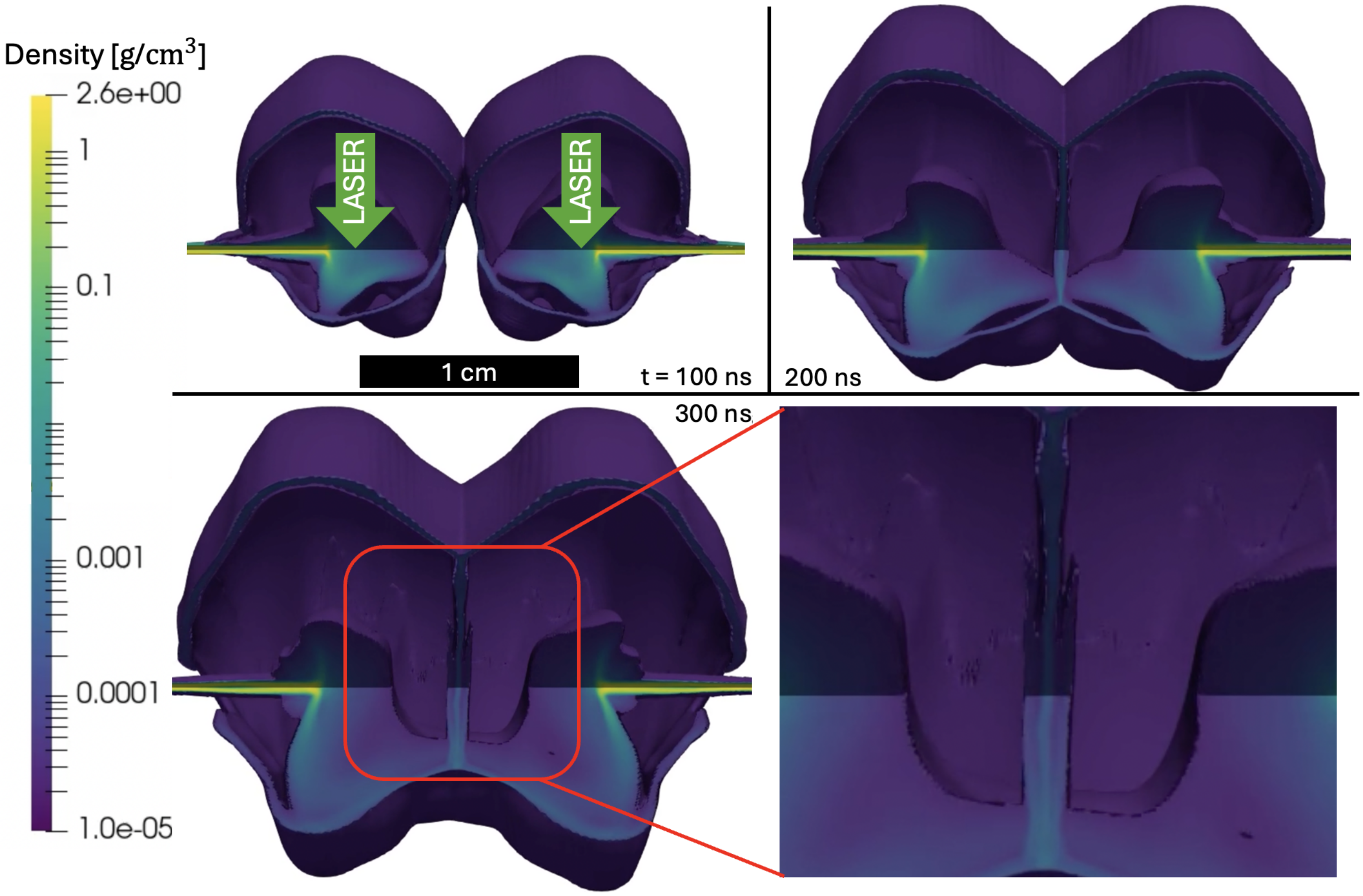

**Figure 6.** Snapshots of the 3D simulation at 100, 200, and 300 ns, showing the collision of two symmetrical blast waves. A quarter of the simulation domain has been removed along the $x$-axis, to reveal the internal structure of the expanding shells. A zoomed-in view at 300 ns provides a clearer visualization of the 3D structure of the interaction region. Densities below $\rho = 1 \times 10^{-5}$ g cm$^{-3}$ were masked to enhance contrast. The locations of the two simultaneous laser pulses impact are indicated by green arrows in the first snapshot.

In summary, this analysis confirms the successful laboratory reproduction of an SNR blast wave and demonstrates that the simulations reliably replicate key experimental observations. These results provide a robust foundation for employing this configuration to investigate SNR collisions under laboratory conditions, while also validating the applicability of the TROLL code for such studies. The numerical work will first focus on the interaction of two simultaneous, symmetrical blast waves, followed by an extension to asymmetrical collisions. Finally, the consequence of a dense obstacle placed along the SNR propagation path will be examined in detail, with particular attention to its influence on both shock dynamics and the response of the obstacle.

### 4.2. Collision of Two Simultaneous SNRs

Two counter-propagating blast waves, analogous to the single blast wave previously described, are generated and ultimately collide, resulting in the formation of an interaction region. The corresponding numerical simulation uses a separation distance of $D_{\rm sphere} = 1$ cm between the two carbon pins (see Figure 2). This collision is illustrated in Figure 6, which presents three snapshots of the simulation at 100, 200, and 300 ns. To better reveal the structure of the interaction region, a zoomed-in view of the 300 ns snapshot is provided in the lower right corner of the figure.

To capture the formation and evolution of this region experimentally, 2D schlieren diagnostics were performed at 150 and 200 ns, shortly after the collision. These diagnostics are numerically reproduced by computing density gradients from the simulation at the same time steps (see Figure 7).

At 150 ns, the reflection of the colliding shocks initiates the formation of the interaction region, highlighted by white arrows. The schlieren visualization markedly enhances the clarity of this structure. The experimental observations reported in Figure 9 of B. Albertazzi et al. (2020) are well-replicated in the simulation. The interaction region forms concurrently in both experiment and simulation, displaying comparable growth by 200 ns. At this juncture, the interaction region's width was measured as 1.5 mm experimentally and quantified as 1.4 mm in the simulation. To more rigorously quantify the formation and evolution of this region, Figure 8 presents the pressure profiles along the $x$-axis from 50 to 300 ns.

At 50 ns, before collision, the two blast waves exhibit similar pressure profiles. Between 100 and 150 ns, the interaction region forms and maintains a relatively uniform

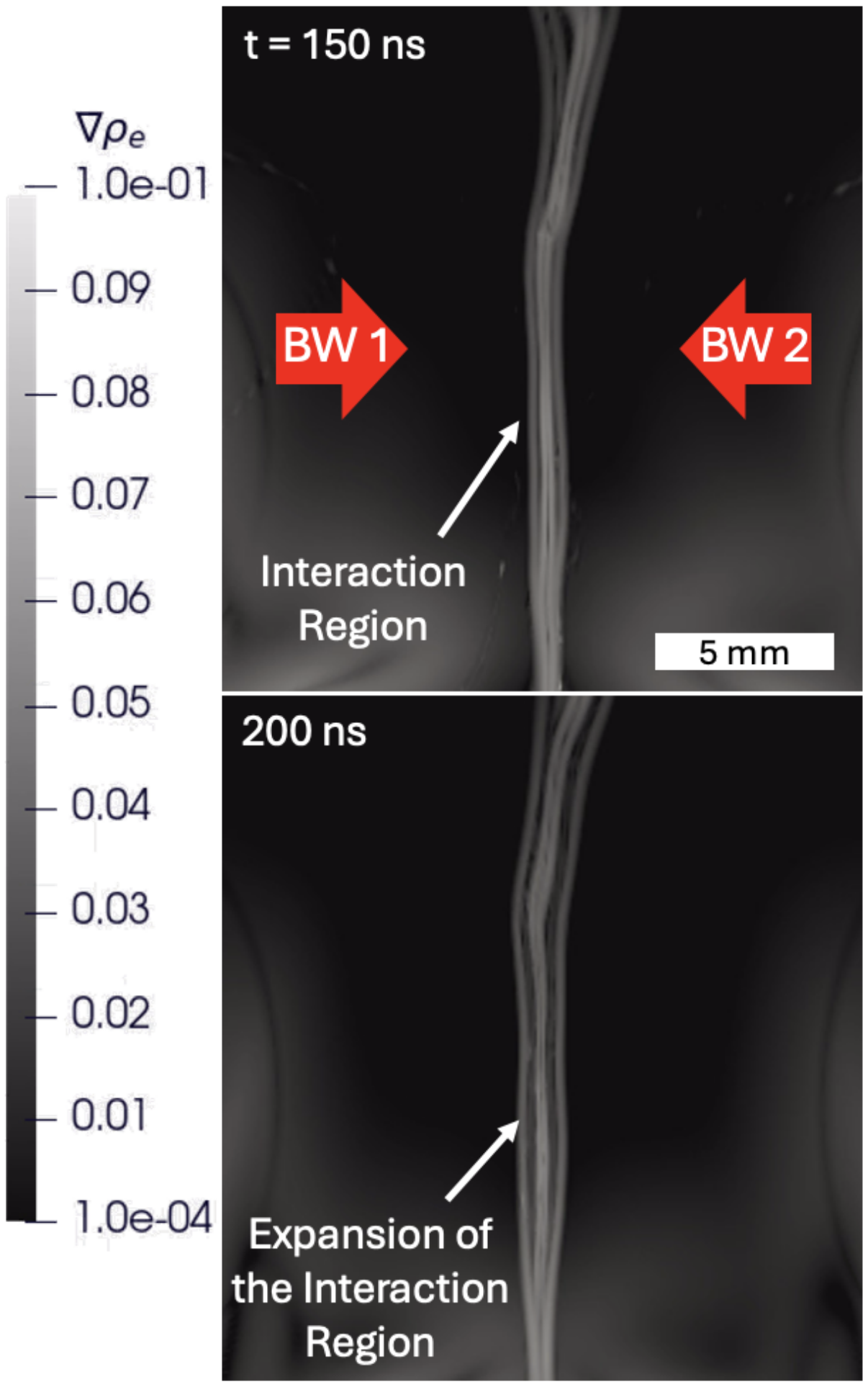

**Figure 7.** The 2D density gradient maps from the 3D simulation, illustrating the formation and evolution of the interaction region at 150 and 200 ns. The interaction region is highlighted with a white arrow for clarity. The flow directions of the counter-propagating blast waves are indicated by red arrows, and a spatial scale is shown in white.

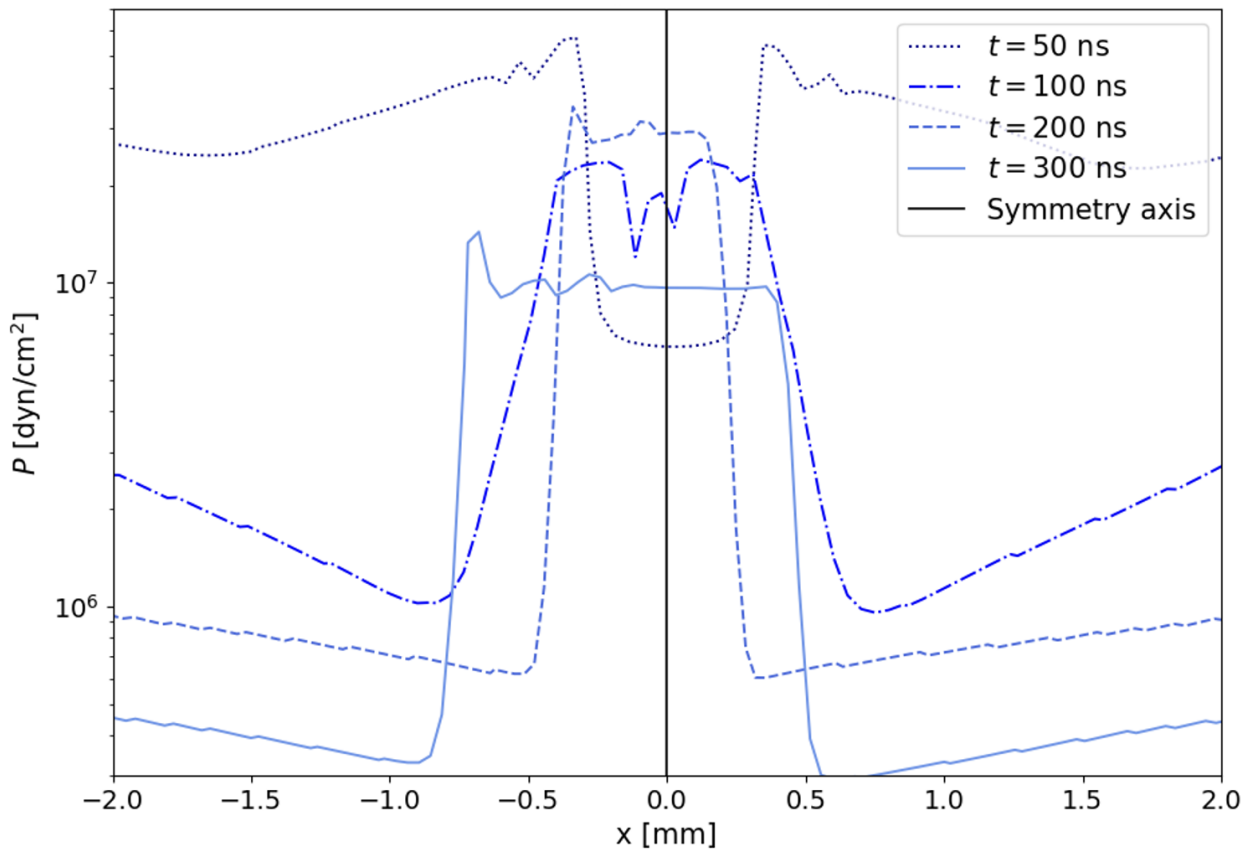

**Figure 8.** Temporal evolution of the pressure profile within the interaction region, extracted along the $x$-axis from 50 to 300 ns. The axis of symmetry is indicated by the black dashed line. Lighter and more continuous blue curves correspond to later times.

pressure distribution, consistent with theoretical expectations. Comparable pressure profiles were obtained numerically by I. V. Krasovskaya & M. P. Syshchikova (1985), although their

**Table 2**
Properties of the Interaction Region Resulting from the Symmetrical Collision of Two Blast Waves

| Parameters | Analytical | Experimental | Numerical |
|---|---|---|---|
| Width (mm) | ⋯ | ∼1.5 | 1.4 |
| Density ratio | 1.6 | ∼1.8 | 2 |
| Pressure ratio | 5–7 | ∼6 | 6.7 |
| Temperature ratio | ⋯ | 1.2 | 1.2 |

**Note.** The analytical values were calculated using the models of R. Y. Tugazakov & A. S. Fonarev (1974) and H. L. Brode (1977), while the experimental values were taken from Section IV of B. Albertazzi et al. (2020). The ratio (e.g., the pressure ratio) is defined as the maximum pressure within the interaction region divided by the pressure of the shocked blast-wave material prior to collision. All properties were evaluated at 200 ns.

peak pressures within the interaction region were lower. This discrepancy likely arises from their one-dimensional geometry, which neglects the obliquity of the colliding shock waves at the interaction region's edges.

In this symmetrical 3D collision, the pressure in the interaction region reaches a maximum value of $P_{\mathrm{IR,max}} \sim 4 \times 10^7$ dyn cm$^{-2}$, approximately 6.7 times greater than the pressure of the material shocked by a single blast wave, $P_1$. This pressure ratio provides a valuable metric for comparison among theoretical models, experimental measurements, and numerical simulations. Similar ratios can be derived for other key physical parameters such as density and temperature. A comprehensive comparison of these ratios, evaluated near 200 ns along with the width of the interaction region, is shown in Table 2.

The strong agreement observed between experimental and numerical data demonstrates the capacity of the TROLL code to accurately replicate this collision in three dimensions, while also showing consistency with theoretical predictions. The pressure ratio values, in particular, corroborate the predictions of H. L. Brode (1977), thereby reaffirming the significant role of the collision geometry in the compression process.

Nonetheless, the interaction region appears more perturbed in the simulation than in the experimental diagnostics, where it exhibits better symmetry (see Figure 9 of B. Albertazzi et al. 2020). This asymmetry manifests as an undulating shape along the upper boundary of the region in Figure 7 and likely arises from subtle deviations from sphericity in the blast waves prior to collision. Across the four time steps in Figure 8, the slight asymmetry initially apparent between the two blast waves' pressure profiles becomes increasingly pronounced. By 300 ns, this asymmetry produces an interaction region that is more extended on one side, measuring approximately 0.9 mm to the left of the symmetry axis and 0.6 mm to the right, reaching an overall width near 1.4 mm. Consequently, the interaction region is roughly 30% more developed in the simulation compared to the experimental observation. This discrepancy is likely due to meshing and ALE resolution effects, since the 3D shock structures are sensitive to these during the calculation. However, these effects do not alter the compression process or the collision geometry along the $x$-axis.

### *4.3. Collision of Two Nonsymmetrical SNRs*

As previously discussed, the experimental setup can be modified to more accurately replicate the DEM L316 system at the laboratory scale. The two SNe forming this system exploded 11,400 yr apart, resulting in an asymmetrical collision. In the laboratory frame, this time delay scales down to approximately 100 ns (see Table 1). The radii of both SNRs were also scaled down to 1 cm, corresponding to a distance of 2 cm between the explosion centers in the laboratory. In this configuration, both blast waves are assumed to carry equal initial energy. Four simulations were performed using the setup shown in Figure 2, with $D_{\rm pin} = 2$ cm, and a second laser pulse fired with delays of $\Delta t = 75$, 100, 125, and 150 ns after the first pulse. These conditions produce blast waves of different ages and radii, leading to an asymmetrical collision. The 3D simulation with $\Delta t = 100$ ns is presented at 100, 200, and 300 ns in Figure 9.

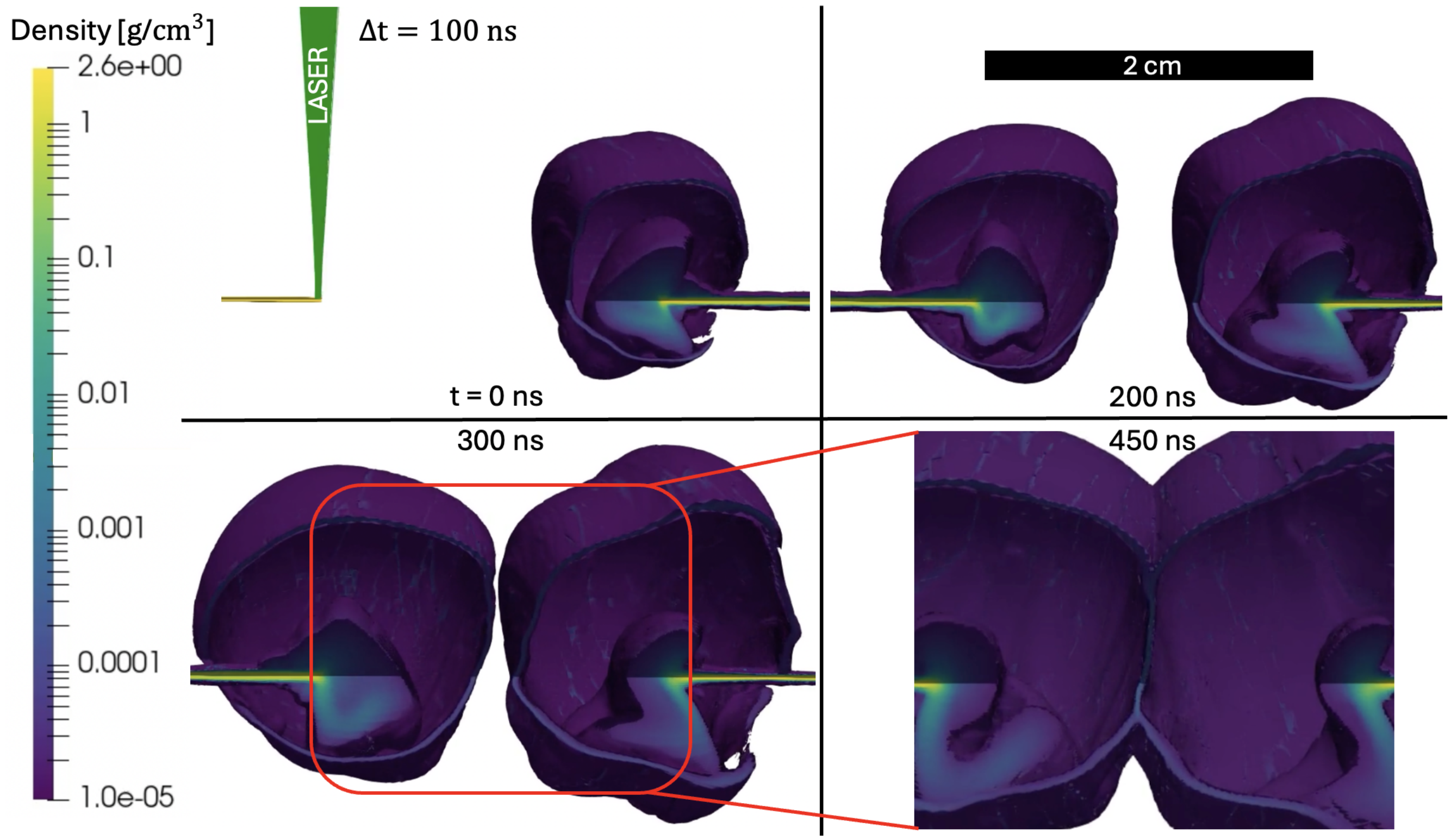

**Figure 9.** Snapshots of the 3D simulation at 100, 200, 300, and 450 ns, showing the collision of two asymmetrical blast waves with a time delay of $\Delta t = 100$ ns. A quarter of the simulation domain along the $x$-axis has been removed to reveal the internal structure of the expanding shells. At 450 ns, a zoom-in to the red square highlights the 3D structure of the interaction zone. Densities below $\rho = 1 \times 10^{-5}$ g cm$^{-3}$ were masked to enhance contrast. The delayed laser pulse is indicated by the green beam in the first snapshot.

The first snapshot at 100 ns illustrates the delay of the second laser pulse, which strikes the second carbon pin while the first blast wave is already well established. At 200 and 300 ns, prior to collision, both blast waves expand asymmetrically in their Sedov–Taylor phase. The first, older blast wave has a larger radius but expands more slowly, exhibiting lower pressure behind its shock front. In contrast, the second, younger blast wave expands slightly faster, resulting in an off-axis collision. The collision occurs around 400 ns, and the resulting interaction region is visible in the lower right panel of Figure 9, which shows a zoomed-in view of the 3D simulation at 450 ns. To characterize this asymmetrical interaction region, the pressure profile along the $x$-axis at the time of collision was analyzed. The results are presented in Figure 10, comparing the pressure distributions for the four different laser pulse delays. The case with no delay, corresponding to a symmetrical collision, is included as a reference.

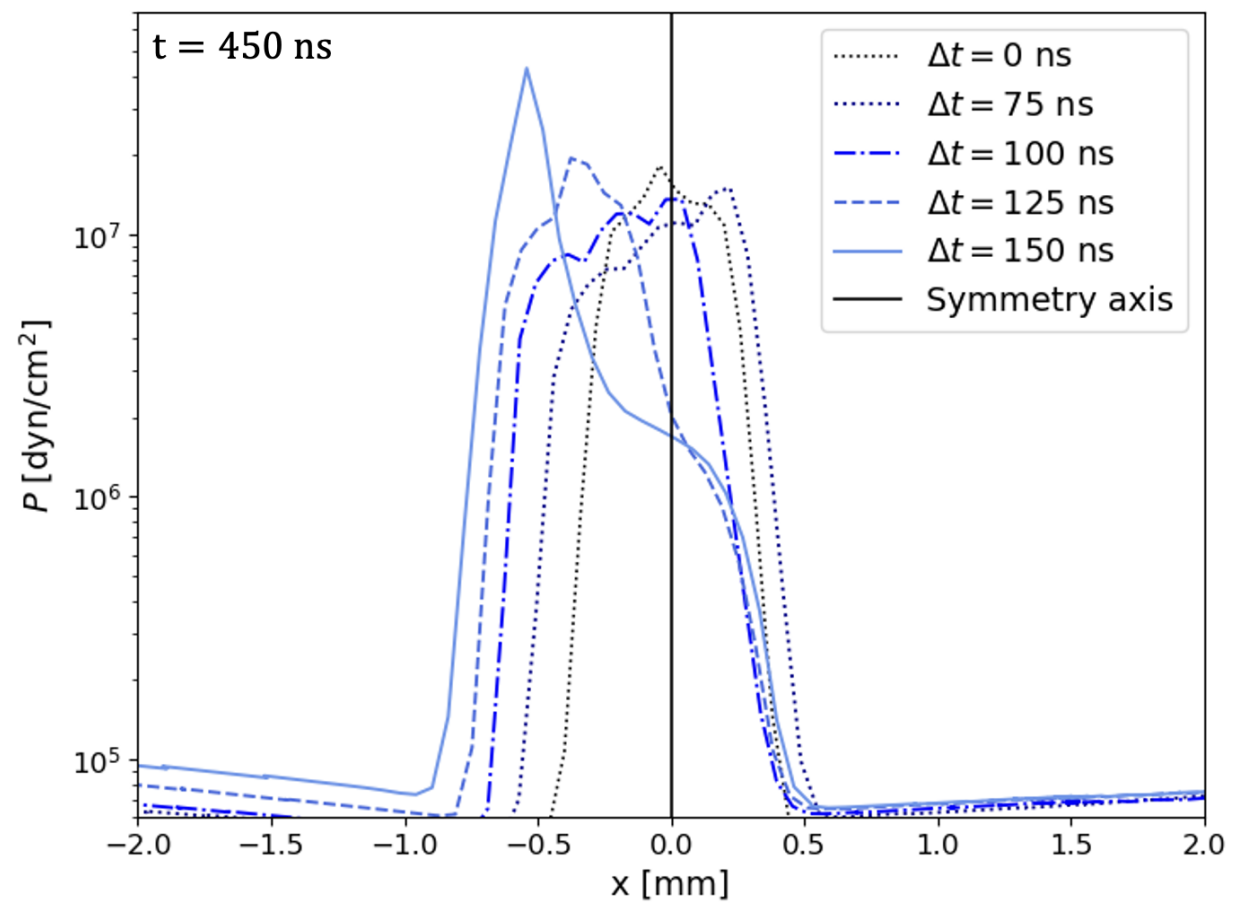

**Figure 10.** Pressure profile of the interaction region along the $x$-axis at 450 ns for the asymmetrical collisions with $\Delta t = 75$–150 ns. The symmetry axis is marked by the black line. The larger the $\Delta t$, the lighter the blue and the more continuous the profile.

As expected from the 3D simulation snapshots, the interaction region shifts away from the symmetry axis ($x = 0$). The greater the delay between the laser pulses, the larger this shift becomes. The darker blue curve, corresponding to $\Delta t = 150$ ns, further illustrates that increasing the delay causes the collision to occur later. This delay affects the formation, properties, and evolution of the interaction region. The region is no longer uniform; its structure becomes asymmetrical and a density gradient is observed, growing over time. However, in this configuration, the impact is not strong enough to produce a clear contact discontinuity, as predicted theoretically (see Section 2.2).

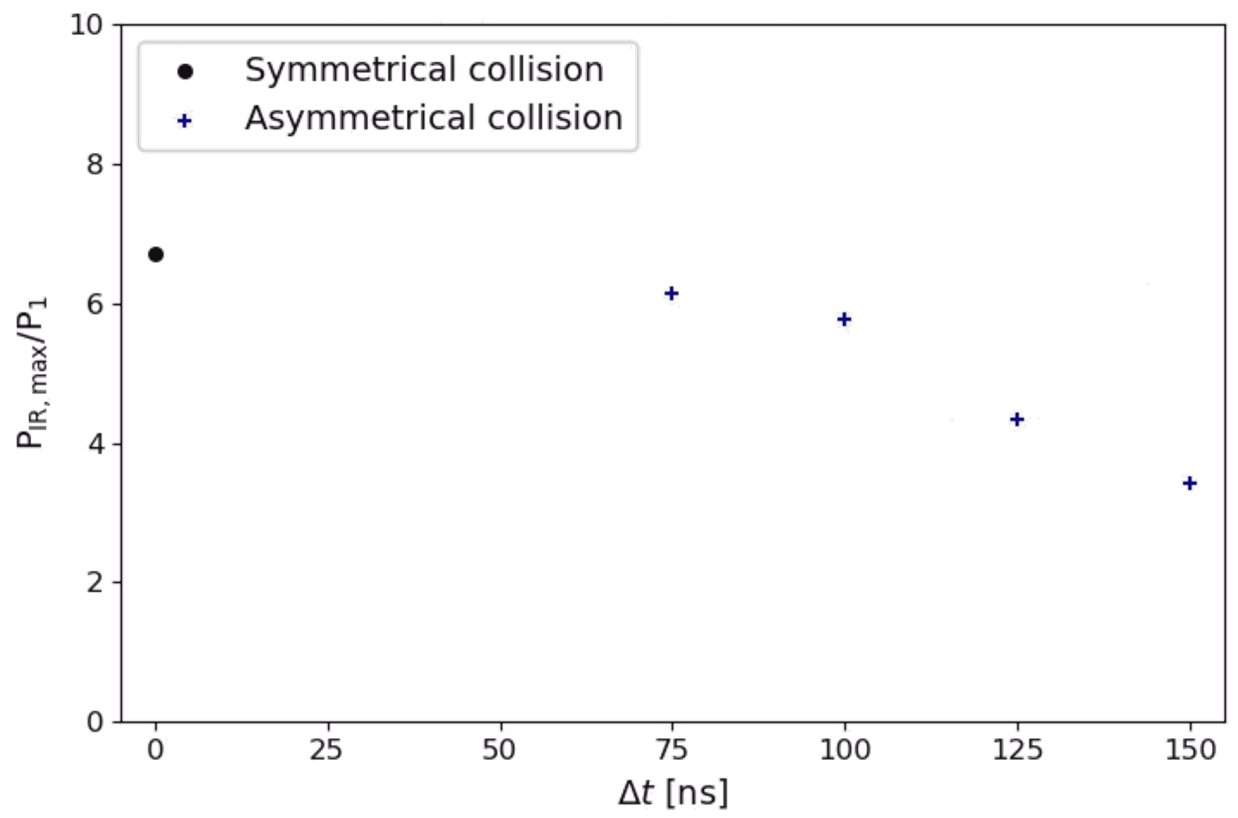

**Figure 11.** Maximum pressure ratio in the interaction region, denoted as $P_{\mathrm{IR,max}}/P_1$, for all laser pulse delay times. The symmetrical case is indicated by the black dot for comparison, while the four asymmetrical cases are shown as blue crosses.

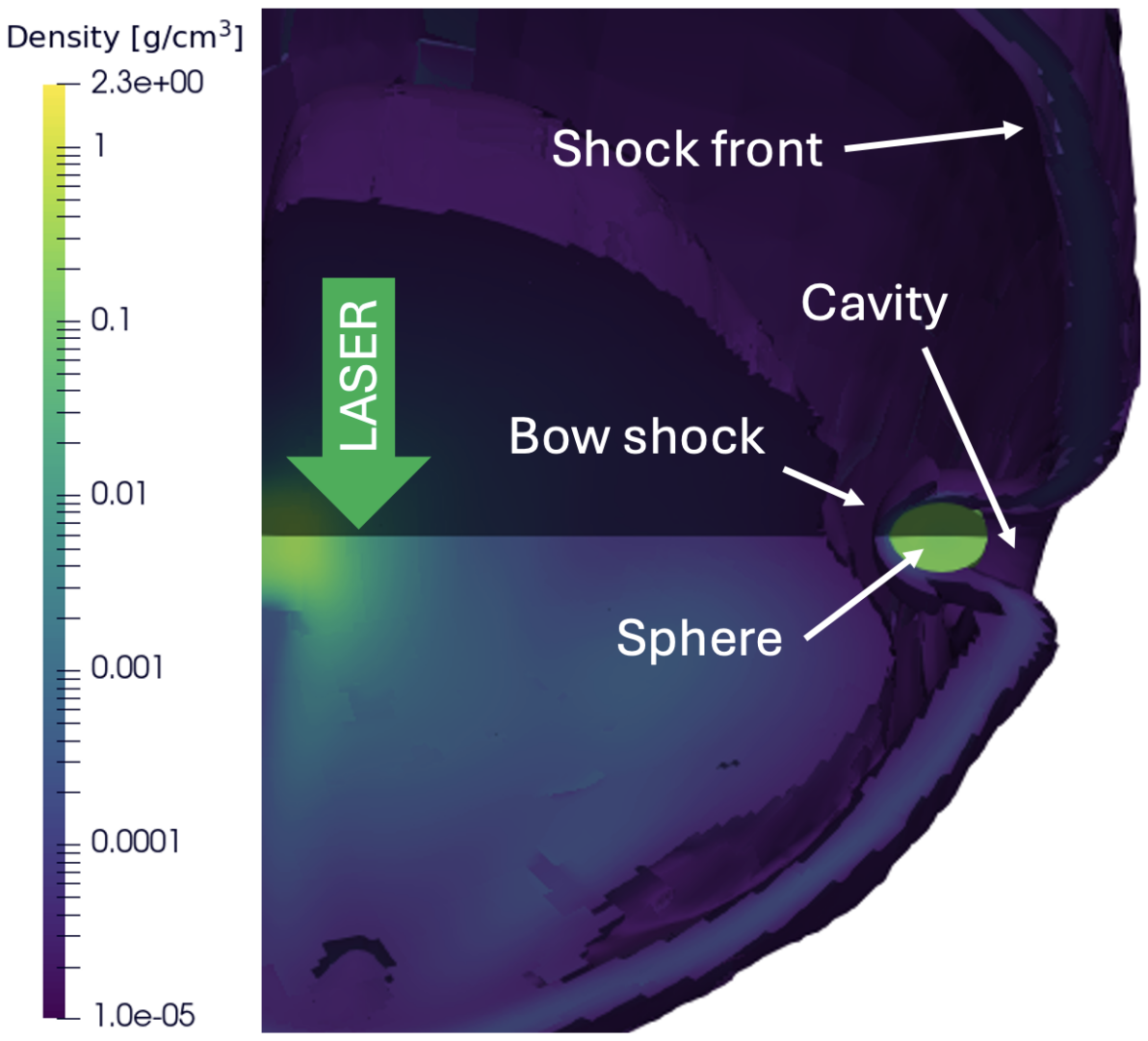

**Figure 12.** Snapshot of the 3D simulation at 200 ns showing the collision of a single blast wave with a dense sphere. A quarter of the simulation domain has been removed along the *x*-axis to reveal the internal structure of the interaction. Densities below $\rho = 1 \times 10^{-5}$ g cm$^{-3}$ were masked to enhance contrast. The shock front, the sphere, the bow shock, and the cavity behind the obstacle are annotated with white arrows.

Figure 10 also shows that the largest delay ($\Delta t = 150$ ns) corresponds to the highest maximum pressure. This is primarily a consequence of visualizing all interaction zones at the same fixed time while the collisions themselves occur at different times. The dark blue curve reaches a higher peak because the collision has just occurred, and the interaction region is not yet fully developed. To quantify this effect, the evolution of the pressure ratio, $P_{\mathrm{IR,max}}/P_1$, was analyzed for all four configurations. This analysis is presented in Figure 11, which also includes the symmetrical case ($\Delta t = 0$ ns) for reference.

This plot demonstrates that increasing asymmetry in the collision leads to reduced compression of the gas within the interaction region. It indicates that a perfectly symmetrical collision maximizes the compression of the material, and any deviation from this configuration must be carefully considered. The maximum pressure decreases from nearly 7 to approximately 4. Furthermore, in this laboratory setup, both blast waves have the same initial energy, which remains constant during expansion. Thus, two nearby SNe releasing equal energy will produce denser interaction regions if their explosions occur simultaneously rather than at different times.

For DEM L316, represented in the laboratory by the $\Delta t = 100$ ns configuration, the pressure in the potential interaction region should be multiplied by approximately 6, resulting in a density ratio (between the interaction region and the shocked material of a single blast wave) slightly below 1.6 and a temperature increase of around 20% (see Table 2). Two-dimensional hydrodynamic simulations of this system at the astrophysical scale were performed by J. C. Toledo-Roy et al. (2009), who found a density ratio of nearly 4 in the interaction region. Their model also included an X-ray emission calculation, expressed as $n^2\Lambda(T)$, where $n$ is the number density and $\Lambda(T)$ is the temperature-dependent radiative cooling function, which enabled an estimate of the expected flux. The predicted flux, about an order of magnitude higher than astronomical observations, led them to suggest that the two SNRs in DEM L316 may not be interacting. The present laboratory-scale study provides new insights into this discussion, predicting a density ratio in the interaction region about twice as low. This reduction decreases the expected emission by a factor of four, bringing it closer to observational values, though it does not definitively confirm the collision scenario. Recent astronomical observations have provided evidence of a star-forming region resulting from the collision of two older SNRs (J. T. Schmelz et al. 2023). This star-forming region, located within the interaction zone of the two SNRs, supports the hypothesis of a collision.

After analyzing the results of both symmetrical and asymmetrical collisions between two blast waves, another configuration relevant to triggered star formation is the impact of SNRs on dense clumps. This scenario is modeled in the laboratory by the interaction of blast waves with solid spheres, which is examined here for various experimental setups.

### 4.4. *Effect of a Dense Clump on the Expansion of SNRs*

The presence of a dense clump along the expansion path of an SNR has two principal and complementary effects. The first is the deformation of the blast wave's morphology. To analyze this phenomenon, a single blast wave was generated, allowing observation of its morphology at later times without interference from a second wave. This deformation is illustrated in 3D in Figure 12, which shows a simulation snapshot at 200 ns, captured shortly after the shock impacts the clump. The blast wave's morphology is visibly perturbed: a cavity forms behind the obstacle, while a bow shock appears upstream. In agreement with the experiment, this deformation is highlighted through the density gradient computed from the 3D simulation as a 2D slice along the symmetry axis (see Figure 13).

These images reveal the cavity-like structure formed as the blast wave wraps around the obstacle and the expansion of the bow shock inside the blast wave's interior, as indicated in the left panel of Figure 13. This collision scenario aligns with findings from A. Y. Poludnenko et al. (2002), who studied interactions between strong shocks and inhomogeneous media, and with experimental diagnostics on Figure 3 of B. Albertazzi et al. (2022). Strong shock waves are stable, and a full reconstruction of the spherical shock front should occur at later times, around 500 ns in this setup. Considering the collision occurs at approximately 100 ns, the blast wave would take about 400 ns to restore its normal morphology in the laboratory frame, corresponding to roughly 25,000 yr at the astrophysical scale based on the scaling in Table 1. This timescale characterizes how long an SNR in the Sedov–Taylor phase takes to recover after a collision, indicating how long such events leave morphological traces. Moreover, the cavity behind the spherical obstacle extends over 2 mm, roughly matching the obstacle's diameter. At the astrophysical scale, this corresponds to a low-density region of about 3 over 1 pc. The right panel of Figure 13 shows that this cavity is subsequently filled with material from the shock surrounding the obstacle, potentially leading to star-forming regions.

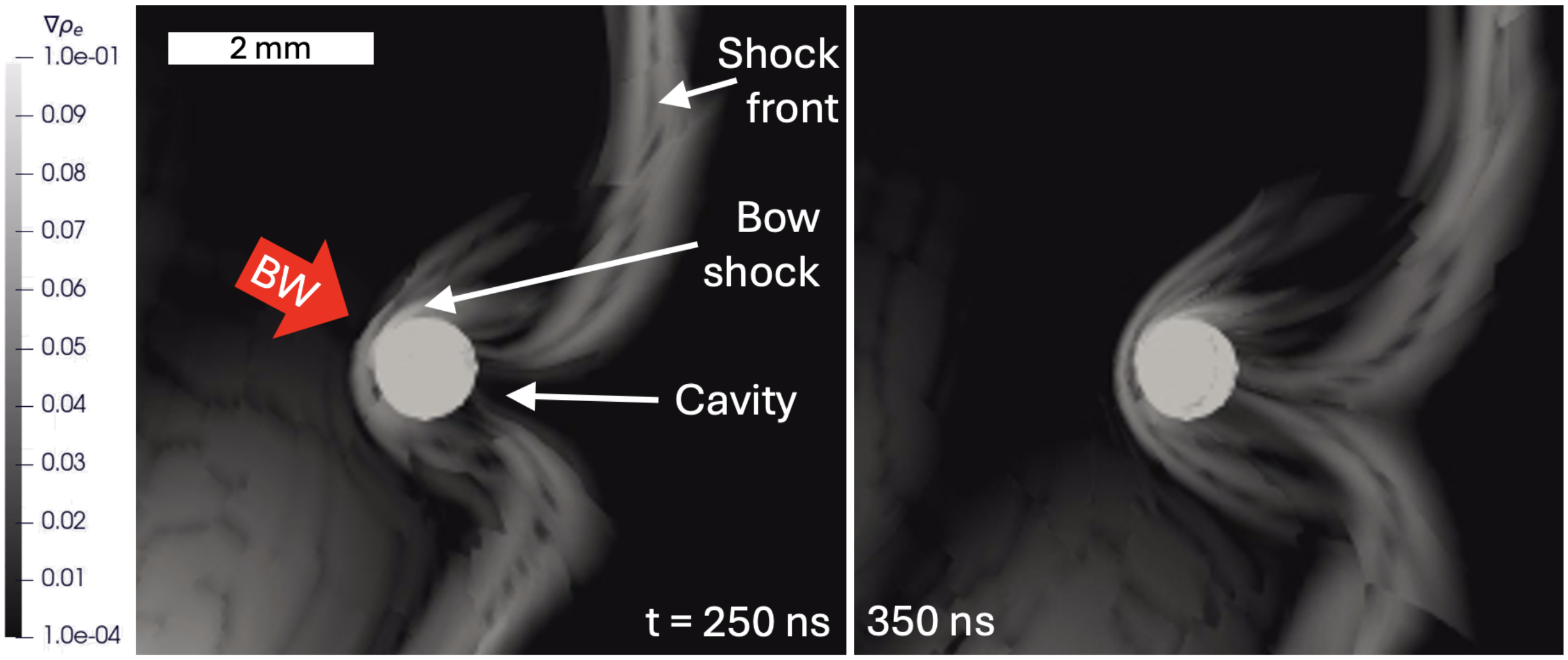

**Figure 13.** The 2D density gradient maps from the simulation at 250 and 350 ns, illustrating the interaction between the blast wave and the sphere. The expansion direction of the blast wave is indicated by the red arrow, and the shock front, bow shock, and cavity are annotated.

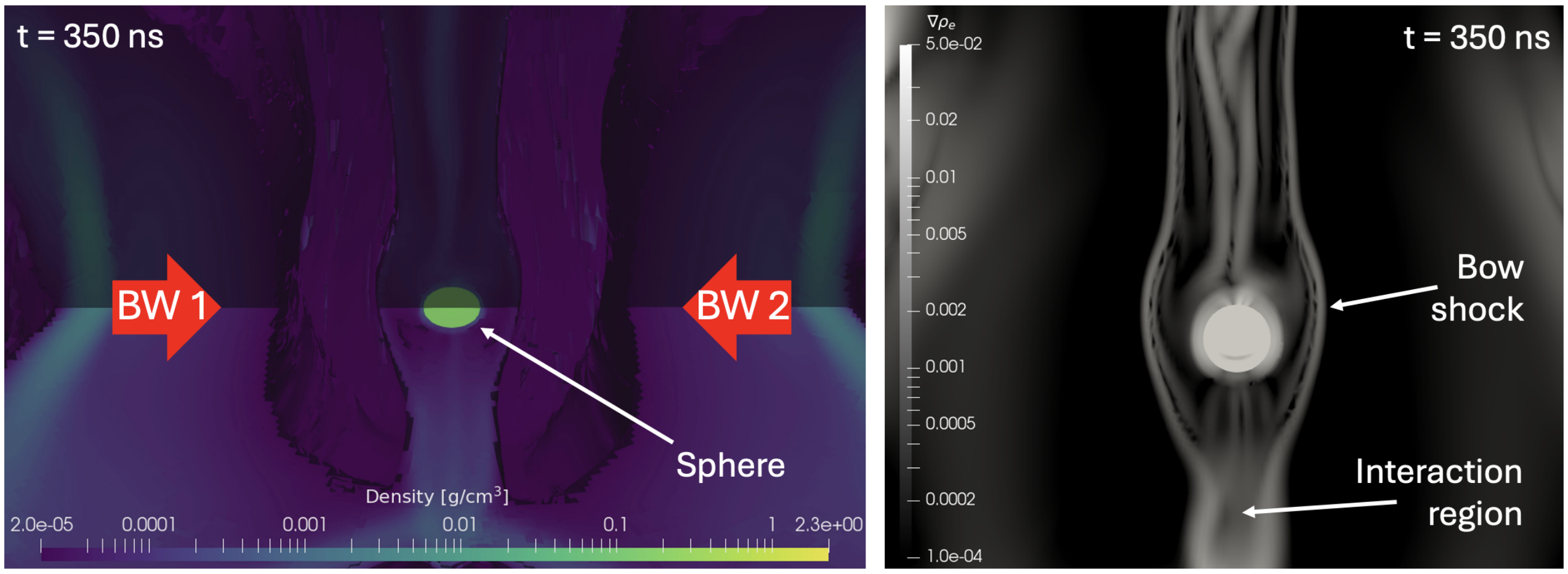

**Figure 14.** (Left) Snapshot of the 3D simulation at 350 ns, showing the collision of two blast waves with the sphere positioned at $D_{sphere} = 2.5$ mm off the $x$-axis. A portion of the simulation volume has been removed to reveal the internal morphology of the interaction. Densities below $\rho = 2 \times 10^{-5}$ g cm$^{-3}$ were masked to enhance contrast. The two blast waves' expansion directions are annotated with the red arrows (BW 1 and 2). (Right) Two-dimensional map of the density gradient from the simulation at the same time, highlighting the interaction between the blast waves and the sphere. The interaction region of the two blast waves and the bow shock surrounding the sphere are annotated in white.

The second effect concerns the compression of the sphere by the blast wave. This begins with the transmission of a shock into the sphere, which then propagates and compresses it. In the experimentally tested setup (with two blast waves), the sphere was placed at $D_{sphere} = 5$ mm off the $x$-axis. Analysis via X-ray radiography showed a reduction in the sphere's volume after impact. This configuration was numerically reproduced in 3D, and simulated X-ray radiographs matched the experimental results well, allowing the study to be extended numerically to new configurations. The simulation also provides additional parameters, such as the sphere's density evolution during the interaction.

The sphere was then moved closer to the $x$-axis, at $D_{sphere} = 2.5$ mm. The 3D simulation snapshot at 350 ns (left panel of Figure 14) illustrates this collision. The results are consistent with the published experimental results (see Figure 6 of B. Albertazzi et al. 2022): the two blast waves strike the sphere diagonally from the top left and top right, causing a similar degree of compression that is slightly enhanced by the sphere's proximity to the blast wave centers. Unlike the single blast wave case, no cavity forms behind the sphere; instead, two bow shocks appear on either side, indicated by white

**Table 3**
Compression Factor of the Sphere Resulting from the Impact of the Two Blast Waves

| $D_{sphere}$ (mm) | Scaled distance (pc) | $C$ | $M_{BE,f}/M_{BE,0}$ |
|---|---|---|---|
| 5 | 11.7 | 1.04 | 0.98 |
| 2.5 | 9.1 | 1.06 | 0.97 |
| −0.5 | 8.1 | 1.1 | 0.95 |

**Note.** The distance from the explosion center to the sphere, respectively, 5, 2.5, and -0.5 mm, is given at the astrophysical scale in the second column. The compression factor is denoted C. The reduction of the Bonnor–Ebert mass resulting from this compression is calculated in the last column.

arrows. These bow shocks disturb the interaction region, which appears broader near the sphere. The density gradient computed from the simulation at the same time (right panel of Figure 14) highlights the influence of the dense obstacle on the propagation of SNR-like blast waves by disrupting the symmetry and the formation of the interaction region. Notably, the maximum pressure reached within the interaction region remains comparable to the case without the sphere. The bow shocks partially deflect and weaken the converging flows, reducing the efficiency of gas compression at the center of the interaction zone.

Finally, placing the sphere directly on the symmetry axis provides further insight into compression mechanisms. Here, the blast waves impact the sphere simultaneously and symmetrically, generating two counter-propagating weak shocks that travel inward, meet at the sphere's center, and reflect off each other. Bow shocks again form on either side, broadening the interaction region and weakening compression. The sphere's diameter decreases by only about 5% compared to its initial size, although the transmitted shocks compress the material to approximately $165\,\mathrm{mg\,cm^{-3}}$. Given its initial density, $\rho_{c,0} = 150\,\mathrm{mg\,cm^{-3}}$, this corresponds to a compression factor of $C = \rho_{c,f}/\rho_{c,0} = 1.1$.

The compression of the sphere in the three configurations with different offsets is summarized in Table 3. This compression allows analysis of the sphere's potential stability. At laboratory scale, gravitational stability can be expressed using the Bonnor–Ebert mass, $M_{BE}$ (C. F. McKee & E. C. Ostriker 2007). Assuming conservation of the sound speed within the clump during compression (i.e., that the transmitted shock is weak), the ratio of this mass before (0) and after ($f$) interaction can be expressed solely in terms of the compression factor $C$ (M. Fontaine et al. 2025):

$$\frac{M_{BE,f}}{M_{BE,0}} = \frac{1}{C^{1/2}}. \qquad (10)$$

This ratio is presented in the last column of Table 3 and illustrates the clump's destabilization by the impact of the two SNRs. Moreover, the closer the clump is to the SNRs' explosion centers, the more compressed it is, especially when its located at the exact center of them, along the symmetry axis.

To extend this work further, laboratory-scale experiments and simulations must be more closely connected to astronomical observables in order to assess how these interactions might be detected. For instance, M. Wardle & F. Yusef-Zadeh (2002) demonstrated that the propagation of SNRs into GMCs could be traced through OH maser emissions, generated by the collision of SN ejecta with fragmented dense regions. Determining the post-collision properties of clumps in the laboratory could provide the necessary physical constraints to model such emissions and enable direct comparisons with astronomical observations.

## 5. Conclusion and Perspectives

The laboratory reproduction of SNRs enables numerous studies of their dynamics and interactions, providing valuable information complementary to astronomical observations. The 3D TROLL simulations presented here have accurately replicated experiments performed on the LULI2000 facility and enabled deeper analysis. Two main scenarios were investigated.

First, the collision of two SNRs leads to the formation of an interaction region whose properties may facilitate star formation. In this region, material compresses by up to a factor of seven, accompanied by an increase in temperature. After reproducing symmetrical collisions consistent with already performed experiments, a detailed numerical study proposed a new configuration to better reproduce the DEM L316 system under laboratory conditions. In this asymmetrical collision case, although the initial explosion energies are identical, any deviation from perfect symmetry reduces material compression and thus decreases efficiency.

Second, the impact of SNRs on dense clumps within GMCs was analyzed. This began with reproducing experimental setups and diagnostics, followed by numerical studies of new configurations varying the clump's position. The results showed two main effects: the obstacle perturbs the SNR morphology by creating a cavity-like structure due to flow diversion; simultaneously, the SNR impact transmits a shock that compresses the clump. The perturbation of the SNR's morphology was detailed, and it was determined that the shock front takes approximately 25,000 yr to reconstruct. The numerical analysis further revealed that the clump's position relative to the SNRs significantly affects compression, with clumps located between two expanding SNRs compressed more effectively by two transmitted shocks. To further advance this work, laboratory-scale experiments and simulations must be connected more closely to astronomical observables, to assess how such interactions could be detected. For example, M. Wardle & F. Yusef-Zadeh (2002) showed that SNR propagation into GMCs could be traced through OH maser emissions generated by collisions of SN ejecta with fragmented dense regions. Determining the post-collision properties of clumps in the laboratory could provide the physical constraints needed to model such emissions and enable direct comparison with astronomical observations.

Future experimental and numerical studies could explore variations such as altering the size and composition of the spheres or generating shock waves in the radiative phase by increasing laser energy and changing the surrounding gas. Although this work focuses on SNRs in their Sedov–Taylor phase, interactions may begin earlier, during the ballistic phase. Explosions within GMCs may lead to multiple SNR–clump interactions, potentially disrupting the environment and contributing to star formation.

## Author Contributions

All authors contributed equally to this paper.

## ORCID iDs

Marin Fontaine https://orcid.org/0009-0000-6974-2692
Clotilde Busschaert https://orcid.org/0000-0002-6975-6581